\documentclass{article}
\usepackage{spconf,amsmath,graphicx,hyperref}
\usepackage{graphicx}
\usepackage{color}
\usepackage{amsfonts}
\usepackage{pifont} 
\usepackage{lineno}
\usepackage{comment}
\usepackage{enumitem} 

\usepackage{kotex}
\usepackage{enumitem}
\usepackage{booktabs,tabularx}
\usepackage{makecell}
\usepackage{multirow}
\usepackage{textcomp}
\usepackage{siunitx}
\usepackage{musicography}
\usepackage{booktabs} 
\usepackage{placeins}

\usepackage{graphicx} 
\usepackage{caption}
\usepackage[normalem]{ulem}

\usepackage{array}       

\title{D3PIA: A Discrete Denoising Diffusion Model for Piano Accompaniment Generation From Lead sheet}
\name{Eunjin Choi, Hounsu Kim, Hayeon Bang, Taegyun Kwon, Juhan Nam}
\address{Graduate School of Culture Technology, KAIST, Republic of Korea}

\begin{document}
\ninept
\maketitle
\begin{abstract}
Generating piano accompaniments in the symbolic music domain is a challenging task that requires producing a complete piece of piano music from given melody and chord constraints, such as those provided by a lead sheet. In this paper, we propose a discrete diffusion-based piano accompaniment generation model, D3PIA, leveraging local alignment between lead sheet and accompaniment in piano-roll representation. D3PIA incorporates Neighborhood Attention (NA) to both encode the lead sheet and condition it for predicting note states in the piano accompaniment. This design enhances local contextual modeling by efficiently attending to nearby melody and chord conditions. We evaluate our model using the POP909 dataset, a widely used benchmark for piano accompaniment generation. Objective evaluation results demonstrate that D3PIA preserves chord conditions more faithfully compared to continuous diffusion–based and Transformer-based baselines. Furthermore, a subjective listening test indicates that D3PIA generates more musically coherent accompaniments than the comparison models.

\end{abstract}
\section{Introduction}\label{sec:introduction}
Accompaniment arrangement is an important music creation process that requires producing accompaniments based on a given melody. This process involves a deep understanding of music theory, stylistic considerations, and structural coherence. 
Among various accompaniment arrangements, piano accompaniment has a long historical significance in Western music (e.g., Liszt's piano arrangement of Lieder) and poses technical challenges due to the wide range of 88 polyphonic keys. Here, we address the task of automatically generating piano accompaniments in the symbolic music domain based on lead sheets where melody and chords are given.

Over the past decade, symbolic music generation has been largely driven by approaches from the natural language processing domain, where music is represented as tokenized sequences of note events and models are trained in a language modeling (LM) fashion. This paradigm has led to the prominence of Transformer-based architectures in numerous studies \cite{huang2018musict, huang2020pop, wu_Compose_2023}. 
While these LM-based approaches have shown impressive results in generating long sequences, they require a complex symbolic music tokenization process. Additionally, their autoregressive generation during inference can lead to error accumulation over long sequences and limit flexibility in controlling the generated samples.

Recently, diffusion models have emerged as powerful tools across domains, including image synthesis \cite{rombach_HighResolution_2022}, audio  \cite{liu_AudioLDM_2024}, and music generation \cite{evans_Longform_2024}, demonstrating excellent capabilities in capturing complex distributions and generating high-quality outputs. 

In symbolic music generation, diffusion models have been applied either directly to piano roll images \cite{min_Polyffusion_2023, wang_WholeSong_2023, huangNoise2MusicTextconditionedMusic2023, zhu_Efficient_2025} or in the latent space of a variational autoencoder (VAE) \cite{mittal_Symbolic_2021, zhang_Composer_2024}. In particular, adapting piano rolls in diffusion models enables more intuitive and simpler representations compared to the LM-based approaches.
While prior work on piano generation with diffusion models used continuous diffusion over piano rolls \cite{min_Polyffusion_2023, wang_WholeSong_2023, zhu_Efficient_2025}, recent findings in transcription show that discrete diffusion handles the inherently discrete (binary on/off) nature of piano rolls effectively \cite{kim_D3RM_2025}. Because diffusion iteratively adds or removes events, the discrete formulation naturally supports arrangement operations (e.g., inserting, deleting, or refining notes/voices). This motivates our investigation into whether discrete diffusion is better suited to symbolic music than continuous approaches.

In this paper, we propose D3PIA, a discrete diffusion-based model specifically designed for the piano accompaniment generation task. Our model consists of a lead sheet encoder and a discrete denoising decoder that operates directly on discrete note states of piano rolls, utilizing dilated NA \cite{hassani2023neighborhood, hassani2022dilated}. The NA-based model architecture effectively captures locally-aligned lead sheet features and their corresponding accompaniments, thereby enhancing the modeling of piano accompaniment according to chord progressions. Through comprehensive comparisons with continuous diffusion and Transformer-based baselines, we show that D3PIA achieves stronger adherence to harmonic constraints and superior performance in subjective evaluations, despite requiring a significantly smaller model size. An ablation study further highlights the contributions of discrete diffusion and our model design to chord coherence and rhythmic consistency. The model training code and generated samples are publicly available \footnote{\url{https://jech2.github.io/D3PIA/}}.

\section{Related Works}\label{sec:related_works}
\subsection{Piano Accompaniment Generation Models}
Since the release of POP909 \cite{pop909-ismir2020}, a dataset of Chinese pop song piano accompaniments, there has been extensive research on generating piano accompaniment from lead sheets \cite{wang_Learning_2020b, zhao_AccoMontage_2021, min_Polyffusion_2023, wang_WholeSong_2023, zhu_Efficient_2025}. Among them, a VAE-based approach \cite{wang_Learning_2020b} generated accompaniments by disentangling chord and texture components, allowing the use of pre-trained chord and texture embeddings to condition generation models \cite{min_Polyffusion_2023, zhao_AccoMontage_2021, wang_WholeSong_2023}. The Accomontage model \cite{zhao_AccoMontage_2021} used template-based retrieval to generate accompaniments.

Polyffusion \cite{min_Polyffusion_2023} and WholeSongGen \cite{wang_WholeSong_2023}, both unconditional symbolic music generation models based on continuous diffusion with piano rolls, can also generate piano accompaniment from lead sheets. Polyffusion incorporates chord conditions using a pre-trained chord encoder \cite{wang_Learning_2020b} with classifier-free guidance. WholeSongGen employs Polyffusion as its backbone in a four-stage cascaded diffusion framework, progressing from song structure generation to piano accompaniment generation. 
FGG \cite{zhu_Efficient_2025}, also based on continuous diffusion, employs chord and onset information as classifier-free guidance for fine-grained conditioning in the piano roll. 

Beyond studies based on POP909, Transformer-based approaches have been applied to piano music generation \cite{huang2018musict, wu_Compose_2023}.
Among them, Compose \& Embellish (C\&E) demonstrated notable performance \cite{wu_Compose_2023} by leveraging two-stage modeling that first generates a lead sheet and then completes the piano accompaniment based on the generated lead sheet. This model adopts the REMI-like representation \cite{huang2020pop}, which interleaves lead sheets and accompaniments at the bar level. 

While these approaches have been successfully adopted in piano accompaniment generation, challenges remain in maintaining coherence with chord conditions, structural consistency, and overall musical quality. In this paper, we employ a discrete diffusion model for piano accompaniment generation and compare it against continuous diffusion-based and Transformer-based models.

\subsection{Discrete Diffusion Models}
Discrete diffusion was first introduced in \cite{sohl-dickstein_Deep_2015a, hoogeboom_Argmax_2021, austin2021structured} by adapting the continuous processes of adding and removing noise to a discrete framework using a transition matrix. This method manipulates system states, with each change governed by specific probabilities. There are a handful of papers that utilized discrete diffusion in symbolic music generation \cite{plasser_Discrete_2023, lv_GETMusic_2023, zhang_Composer_2024}. They applied discrete diffusion on the latent space of VQ-VAE \cite{zhang_Composer_2024}, on the multi-hot music event token vector per each time step \cite{plasser_Discrete_2023}, and on the compound tokens of pitch and duration for each note \cite{lv_GETMusic_2023}. To apply discrete diffusion to piano roll representations, one can either treat the piano roll as an image and operate in the VQ-VAE latent space \cite{zhang_Composer_2024} or define note states for each piano roll pixel and denoise them individually \cite{kim_D3RM_2025}. The latter approach, named D3RM, has proven effective in piano transcription by applying the diffusion process to discrete note states in piano rolls.
Their use of stacked NA in the denoising module enabled efficient handling of high-resolution piano rolls while aligning local audio features with the transcribed notes. In this paper, we adopt D3RM as our backbone, leveraging its proven effectiveness in locality-aware attention and discrete-state prediction.

\begin{figure}[h]
 \centering
 \includegraphics[width=0.7\columnwidth]{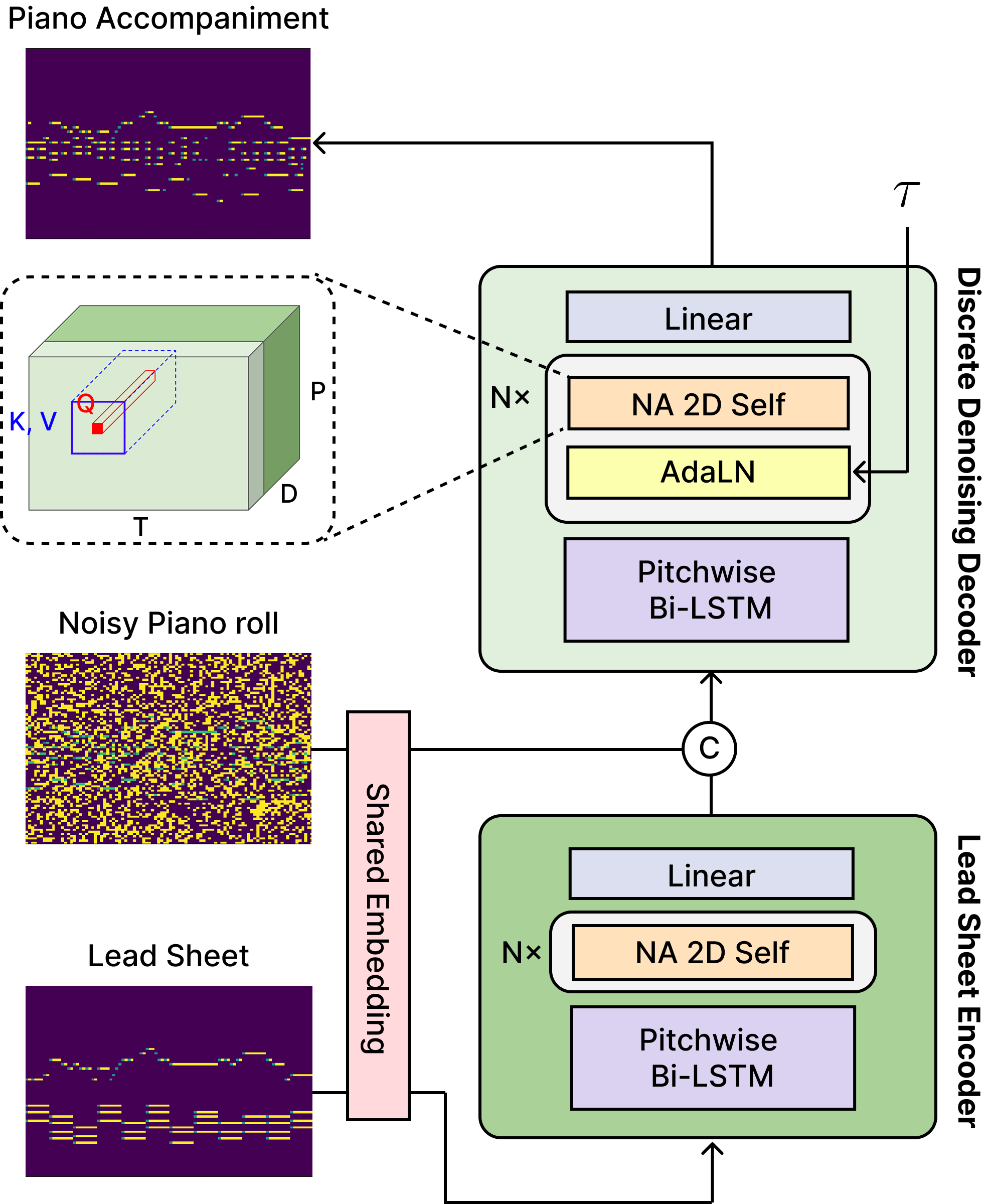}
 \caption{D3PIA model structure. Lead sheet information is attended to in the encoder, and the encoder output (shown in dark green) is concatenated (denoted as C) with the noisy piano roll and further attended to in the denoising decoder via NA. N denotes the number of NA layers. In the NA 2D Self module, T, D, and P refer to timestep, embedding dimension, and pitch, respectively, while Q, K, and V represent the query, key, and value used for attention computation. $\tau$ indicates the diffusion timestep.}
\vspace{-10pt}

\label{fig:model_structure}
\end{figure}

\section{Methods}\label{sec:dataset_representation}

\subsection{Data Representation}
The piano roll representation effectively captures local relationships in both the vertical (harmony) and horizontal (rhythm) dimensions. Following previous works \cite{min_Polyffusion_2023, wang_WholeSong_2023, zhu_Efficient_2025}, we represent the piano roll using a 16th-note time unit. For pitch, we use 88 values corresponding to MIDI notes from 21 to 108. We label each piano roll value with one of four states: \textit{onset}, \textit{off}, \textit{sustain}, and \textit{MASK}. For the lead sheet, melody and chord labels are combined in a single piano roll channel by converting all chord labels into MIDI pitches starting from C4 and shifting them 2 octaves lower, preventing overlap with melody.

\subsection{Model Architecture}\label{subsec:model architecture}
Our proposed model, D3PIA, consists of an encoder and a denoising decoder, as shown in Figure \ref{fig:model_structure}. The encoder receives the lead sheet piano roll as input, and its output features are concatenated with the noisy accompaniment piano roll along the channel dimension before being passed to the denoising decoder.
Both the encoder and decoder include a pitch-wise bidirectional LSTM layer to capture temporal transitions within each pitch, as well as NA 2D self-attention blocks followed by a linear layer. The primary difference is that the decoder incorporates Adaptive Layer Normalization (AdaLN) layers for conditioning diffusion timesteps. The denoising decoder's embedding layer is shared with the encoder, leveraging the equivalent piano roll structure of both the noisy input and lead sheet condition. 
Using the NA 2D encoder for the lead sheet condition effectively captures the local vertical relationship between chord progressions and melody while preserving its piano roll shape. During training, the model receives both the noisy input accompaniment and the lead sheet condition; at inference, only the lead sheet condition is provided, and the model generates the accompaniment via iterative denoising.

\subsection{Discrete Diffusion for Piano accompaniment}
%
We converted each piano roll element into a one-hot vector, resulting in $\mathbf{y} \in \mathbb{R}^{T \times 88 \times 4}$, where $T$ is the time frame length, 88 are MIDI notes, and 4 is the number of the note states mentioned above.
The forward process consists of gradually masking or perturbing the states of each element of the piano roll, by multiplying the following transition matrix $[\mathbf{Q}_\tau]_{ij}=q(y_\tau^{tp}=j|y_{\tau-1}^{tp}=i)\in\mathbb{R}^{\;4\times4}$ to the final dimension,
\begin{equation}\label{eq:1}
    \mathbf{Q}_\tau =
    \begin{bmatrix}
    \alpha_\tau+\beta_\tau & \beta_\tau & \beta_\tau & 0\\
    \beta_\tau & \alpha_\tau+\beta_\tau & \beta_\tau & 0\\
    \beta_\tau & \beta_\tau & \alpha_\tau+\beta_\tau & 0\\
    \gamma_\tau & \gamma_\tau & \gamma_\tau & 1\\
    \end{bmatrix}
\end{equation}
where $y_{\tau}^{tp}$ denotes the state of a label $\mathbf{y}$ at time position $t\in \{0, 1, ..., T-1\}$, pitch $p \in \{0, 1, ..., 87\}$ and diffusion timestep $\tau \in \{0, 1, ..., \mathcal{T}-1\}$; $\alpha_\tau$, $\beta_\tau$, and $\gamma_\tau$ are the probabilities of preserving, perturbing, and masking the state, respectively. The backward process is predicting the reverse probability distribution $p_\theta(\mathbf{y}_{\tau-1}|\mathbf{y}_\tau, \mathbf{x})$, where $\mathbf{x}$ is the piano lead sheet represented as the same shape, $\mathbf{x} \in \mathbb{R}^{T \times 88 \times 4}$. The training objective is described as follows:

\vspace{-5pt}

\begin{equation}
\begin{split}
    \mathcal{L}_{vlb} &= D_{KL}\big[q(\mathbf{y}_\mathcal{T}|\mathbf{y}_0)||p(\mathbf{y}_\mathcal{T})\big] \\& + \sum^{\mathcal{T}}_{\tau=1}\big[D_{KL}[q(\mathbf{y}_{\tau-1}|\mathbf{y}_{\tau},\mathbf{y}_0)||p_{\theta}(\mathbf{y}_{\tau-1}|\mathbf{y}_{\tau},\mathbf{x})]\big]
\end{split}
\end{equation}

We additionally implemented the absorbing-state (AS) sampling method from \cite{kim_D3RM_2025}, which has been proven to enhance the refinement capability of discrete diffusion models for the piano transcription task—a discriminative task requiring extremely high precision. The model is trained to reverse the perturbation and masking according to the schedule described in Section \ref{subsec:model configuration}, and during inference, it generates samples with $\beta_t=0$ throughout all diffusion timesteps.

\section{Experiment}\label{sec:experiment}

\subsection{Dataset}
We used POP909 \cite{pop909-ismir2020}, which is a widely used benchmark for piano accompaniment generation consisting of 909 MIDI files.
We split the dataset into an 8:1:1 train:valid:test ratio, and refer to the test set as POP909-test for convenience.
For training, we randomly cropped segments to 8-bar lengths and applied random pitch transposition from -5 to 6 semitones for augmentation.  
For evaluation, we utilized the entire POP909-test comprising 86 pieces. To ensure fair comparison between our diffusion-based models (capable of generating 8 bar segments) and the Transformer-based model, we segmented the POP909-test into 8 bar segments, yielding a total of 905 segments.

\subsection{Model Configurations}\label{subsec:model configuration}

Each piano-roll state was represented as a 4-dimensional embedding vector. Unlike the original D3RM model, which was designed for the piano transcription task, D3PIA is a generative model that requires greater model capacity, with the ability to perceive larger input contexts and to be trained on a larger dataset. Therefore, we scaled up the model by increasing the layer size of the decoder and the window size of dilated NA in each layer to create a larger receptive field. Based on our preliminary study, we set the number of layers to 10, and the dilated window size to 5 with a dilation ratio of [1, 2, 4, 8, 16, 1, 2, 4, 8, 16]. The model was trained with a batch size of 8 using a single NVIDIA A6000 GPU. Training was run for 200k steps, after which the note-level F1 score on the validation set stabilized. For the diffusion process, we used $T = 100$ timesteps. At each step, $\alpha_t$, $\gamma_t$, and $\beta_t$ was scheduled following \cite{gu2022vector, kim_D3RM_2025}. The auxiliary loss weight was set to $\lambda = 5.0 \times 10^{-4}$. Optimization was performed using AdamW with $\beta=(0.9, 0.96)$. The initial learning rate of $1.0 \times 10^{-3}$ was reduced by a factor of 0.8 whenever the diffusion loss did not improve for 25k steps.

\subsection {Comparison Models} \label{subsec:comparison_models}
We compared the model performance with diffusion-based models and a Transformer-based model that accepts both melody and chord conditions. For the diffusion-based models, we selected Polyffusion \cite{min_Polyffusion_2023}, 4th stage model of WholeSongGen (WSG-4th) \cite{wang_WholeSong_2023}, and FGG \cite{zhu_Efficient_2025} for comparison. For the Transformer-based model, we selected the Embellish part of C\&E (C\&E-E). Accomontage \cite{zhao_AccoMontage_2021} was excluded from our comparisons because it was trained on a different POP909 split, and no official training code is available. Also, we could not compare our model with GETMusic \cite{lv_GETMusic_2023}, as it is a large-scale model 
with checkpoints unavailable. 

For C\&E-E, we reproduced the model and trained it without velocity tokens. For all other models, we used official code with minimal modifications to enable as fair a comparison as possible; For FGG, we modified the original 4 bar length input to 8 bars. We identify Polyffusion and C\&E-E as our direct comparison model and note that we cannot fairly compare our model with all comparison models, since some models require extra input in addition to melody and chord, as shown in Table \ref{tab:model structure comparison}.

 
Since the POP909 dataset does not provide an official split, existing models use different random splits. Therefore, we retrained all models using our split. All models were trained with random pitch transposition augmentation. We confirmed that the re-trained models achieved similar generation quality to the official pre-trained checkpoints, if provided, based on evaluation metrics.

\begin{table}[t]
\centering
\small
\resizebox{\columnwidth}{!}{%

\begin{tabular}{c|c|c}
\toprule
\textbf{Model} & \textbf{Base} & \textbf{Model Input} \\
\midrule
Polyffusion \cite{min_Polyffusion_2023} & C & melody, chord (encoded chroma vector \cite{wang_Learning_2020b}) \\
WSG-4th \cite{wang_WholeSong_2023} & C & melody, chord (piano roll), song structure, reduced lead sheet \\
FGG \cite{zhu_Efficient_2025} & C & melody, chord (piano roll; expanded to whole octave), onset
\\
C\&E-E \cite{wu_Compose_2023} & T & melody, chord (labels as tokens) \\
D3PIA (Ours) & D & melody, chord (piano roll) \\
\bottomrule
\end{tabular}
}
\caption{Comparison of all methods in terms of model architecture, inputs used in experiments. C, D refers to continuous and discrete diffusion and T represent Transformer.}

\vspace{-10pt}
\label{tab:model structure comparison}
\end{table}

\subsection{Evaluation Metrics}\label{subsec:evaluation_metrics}
\subsubsection{Objective Evaluation}
As discussed in Section \ref{sec:introduction}, the primary goal of accompaniment is to follow the given melody and chord conditions while maintaining a consistent rhythm pattern. Considering this, we assessed the harmonic coherence and rhythmic consistency of generated accompaniments. To evaluate harmonic coherence, we used the out-of-key tone ratio (\textbf{OOK}) \cite{zhu_Efficient_2025}, chord accuracy (\textbf{CA}) \cite{ren2020popmag, zhu_Efficient_2025}, and chord similarity (\textbf{CS}) \cite{zhu_Efficient_2025}. OOK was calculated by counting the number of out-of-key notes based on the POP909 key-signature labels and normalizing the count. For CA and CS, chords are first extracted from generated samples with the rule-based method of Dai et al. \cite{dai2020automatic} and compared with POP909-test references: CA counts exact beat-level matches, whereas CS measures cosine similarity between two-bar chord embeddings from a pre-trained chord encoder \cite{wang_Learning_2020b}, thus crediting overlaps in chord tones. Because CA cannot capture such overlaps, we additionally report CS. For rhythmic consistency, we use grooving pattern similarity (\textbf{GS}) \cite{wu_Jazz_2020}, which measures pairwise similarity between per-bar binary onset vectors within a segment.

\subsubsection{Subjective Evaluation}
We conducted a listening test to assess the musical quality. From POP909-test, we randomly selected 10 8-bar excerpts and generated accompaniments from each model given the corresponding lead sheets. All stimuli were anonymized and presented in randomized order. Thirteen participants (each with at least a Bachelor’s degree in piano or composition) provided mean opinion scores on a 5-point Likert scale. The listening criteria are as follows:

\begin{description}[leftmargin=1.7em, labelsep=0.4em, itemsep=0pt, topsep=2pt]
\item[\textbf{Coherence}] Coheres with the given melody?
\item[\textbf{Harmony}] Follows chord progression; no missing chord tones?
\item[\textbf{Consistency}] Rhythmic consistency maintained?
\item[\textbf{Correctness}] Any inharmonious notes, unnatural rhythm, or awkward phrasing?
\item[\textbf{Overall}] Overall musical quality?
\end{description}

\begin{table}[t]
\centering
\resizebox{.98\columnwidth}{!}{%
\small 
\setlength{\tabcolsep}{4pt} 
\begin{tabular}{ccccccc}
\toprule
\multirow{2}{*}{\textbf{Model}} & \multirow{2}{*}{\textbf{\# Params}} & \multirow{2}{*}[-0.25em]{\makecell{\textbf{Inf. Time} \\ (sec)}} & \multicolumn{3}{c}{\textbf{Harmony}}         & \textbf{Rhythm} \\ \cmidrule{4-7}
                                 &                &                     & OOK (\%)     & CA (\%)       & CS (\%)       & GS (\%)         \\ \midrule
GT      & -          & -                                   & \textbf{0.0} & 91.6          & 95.7          & 82.7            \\ \midrule
Polyffusion       & 41.1M               & 21.4                               & \textbf{0.0} & 37.5          & 54.0          & 79.9            \\
C\&E-E        & 66.0M                   & 18.7                               & 14.8         & 58.1          & 70.6          & \uline{80.8}            \\
\textbf{D3PIA}    & \textbf{2.2M}         & \uline{1.7}                       & \textbf{0.0} & \uline{80.1}          & \uline{93.6}          & \textbf{82.1} \\
\midrule
WSG-4th*  & 41.6M                          & 79.0                              & 2.4          & \textbf{87.6} & \textbf{94.6} & 75.4            \\
FGG*           & 36.7M                   & \textbf{0.4}                               & \textbf{0.0} & 62.0          & 77.3          & 78.9         \\ \bottomrule
\end{tabular}
}
\vspace{-2pt}
\caption{Objective results on POP909-test. OOK (out-of-key ratio, lower), CA (chord accuracy, higher), CS (chord similarity, higher), GS (grooving similarity, closer to GT), OA (overlapping area, higher). Best in \textbf{bold}, second best \underline{underlined}. * indicates the model incorporates extra inputs, as shown in Table ~\ref{tab:model structure comparison}. Inference time is reported for one 8-bar sample.}
\vspace{-3pt}

\label{tab:Objective Evaluation ALL}
\end{table}

\begin{figure}
  \includegraphics[width=\columnwidth]{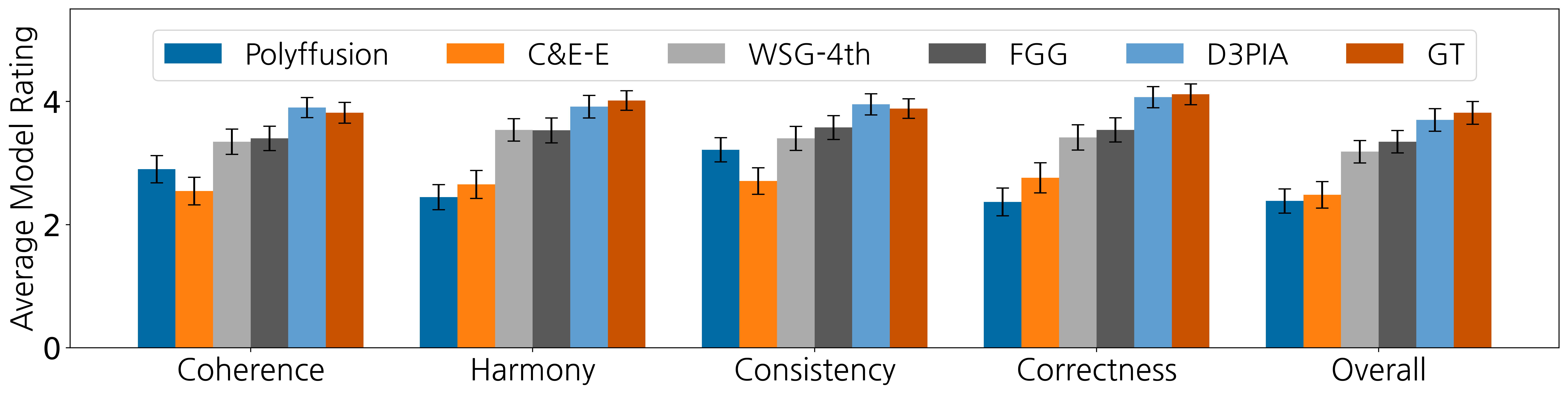}
 \caption{Subjective evaluation result with 95 \% confidence interval.}
\vspace{-15pt}
\label{fig:subjective evaluation}
\end{figure}

\vspace{-5pt}

\section{Results \& Discussions}
\vspace{-4pt}
\subsection{Comparative Results}
\label{subsec:comparative results}
Table \ref{tab:Objective Evaluation ALL} shows the objective results on the POP909-test. D3PIA and WSG-4 achieved the highest CA and CS metrics and comparable CS values to those of GT. For C\&E-E, their chord condition coherence was lower but comparable to other diffusion-based models, suggesting that the Transformer-based model can predict chord pitches to some extent with minimal chord label information. However, they produced frequent out-of-key notes. 
For rhythm, D3PIA achieved a higher GS than other models, indicating more consistent rhythmic patterns than other models. In previous music generation studies \cite{wu_Jazz_2020, wu_Compose_2023}, GS values closer to the reference set were considered better. C\&E-E achieved the second-best GS value, which was expected, since Transformer-based models have long receptive ranges that support rhythmic coherence. 

In the subjective listening test, as shown in Fig. \ref{fig:subjective evaluation}, D3PIA outperformed all comparison models, including the main comparison model, Polyffusion and C\&E-E. A notable result is the performance of WSG-4th, which achieved higher scores than D3PIA on CA and CS but not on the Harmony metric in subjective evaluation. Further analysis of the listening test samples revealed that several WSG-4th outputs received Harmony scores around 3, despite very high CS values ($>$0.95) \footnote{We invite readers to check these examples in the supplementary webpage: \url{https://jech2.github.io/D3PIA/}}.
We attribute this discrepancy to perceived dissonant notes produced by WSG-4th, consistent with the OOK results in Table~\ref{tab:Objective Evaluation ALL}. Our subjective evaluation accounted for both coherence and correctness, meaning such out-of-key mistakes could have a significant impact on the scores.

Additionally, we report the inference time for generating one 8-bar sample in Table~\ref{tab:Objective Evaluation ALL}. D3PIA achieves substantially faster inference (1.7 sec) than other models, enabling real-time generation, while remaining slower than FGG, which applies DDIM sampling.

\subsection{Ablation Study}
\label{subsec:ablation}
When we listened to generated samples, we observed that D3PIA produces dissonant notes less frequently than other models. We conjecture that this improvement stems from the use of the NA-based lead sheet encoder. Another possible explanation is the note refinement capability of the discrete diffusion model, as discussed in \cite{kim_D3RM_2025}. To investigate these, we conducted an ablation study (Table \ref{tab:ablation}).

First, we compared models where the lead sheet was directly concatenated to the original D3RM decoder without the use of an encoder. The results show that utilizing the original D3RM decoder (only 477K parameters) achieved a relatively high CA and CS values around 80\% with no out-of-key tone, still outperforming Polyffusion and C\&E-E. This indicates the effectiveness of the discrete diffusion architecture. Moreover, additional GS gain was observed after scaling the model up to the D3PIA configuration. However, these models performed worse than D3PIA in terms of harmonic coherence and rhythmic consistency, highlighting the importance of the proposed NA-based lead sheet encoder. Next, excluding chords from the encoder caused CS to drop drastically (to 59.2\%), demonstrating the effectiveness of the chord input conditioning in the encoder. Furthermore, removing AS sampling reduced CS by about 16.4\% compared to full D3PIA. This implies that the refinement process in the discrete diffusion model (via AS sampling) corrects the note states, generating more coherent accompaniments given the lead sheet. 

\begin{table}[t]
\centering
\small 
\resizebox{0.85\columnwidth}{!}{%

\begin{tabular}{cccc}
\toprule
\multirow{2}{*}{\textbf{Model}} & \multicolumn{2}{c}{\textbf{Harmony}} & \textbf{Rhythm} \\ \cmidrule{2-4}
                                 & CA (\%)           & CS (\%)          & GS (\%)         \\ \midrule
original D3RM decoder            & 79.9              & 78.0             & 74.9            \\
D3RM decoder w/ scale-up         & 79.8              & 78.0             & 77.1            \\
\midrule
D3PIA w/o chord in encoder       & 35.9              & 59.2             & 83.9            \\
D3PIA w/o AS sampling            & 76.8              & 77.2             & 79.6            \\
\textbf{D3PIA}                   & \textbf{80.1}     & \textbf{93.6}    & \textbf{82.1} \\ \bottomrule  
\end{tabular}
}
\caption{Ablation study of D3PIA model. All had 0\% OOK.}
\vspace{-18pt}

\label{tab:ablation}
\end{table}

\section{Conclusion}
Unlike general music generation tasks that often prioritize creativity and diversity, accompaniment generation critically depends on reliably following melody and chord progression. In this work, we highlighted two main take-away messages. First, modeling at the discrete note-state level is highly effective for symbolic music generation. Second, generating accompaniment from a given melody and chord progression requires strong local dependency, which should be explicitly reflected in the model design. While strict adherence to chord tones may reduce diversity, it prevents perceptually dissonant notes and enhances controllability, which are crucial for practical accompaniment systems.

A key limitation of D3PIA lies in its discretized note representation, which does not model velocity. Future research may extend the note state space or incorporate velocity prediction models \cite{kim_Method_2024}. 
Another promising direction is to generate more diverse and artistically rich accompaniments while maintaining adherence to the given conditions. This could be achieved by scaling up the model and input length and by enhancing controllability through style variation with pre-trained texture embeddings \cite{wang_Learning_2020b}.

\vfill\pagebreak

\section{Acknowledgments}
\vspace{-5pt}
This work has been supported by the National Research Foundation of Korea (NRF) grant funded by the Korea government (MSIT) under Grant RS-2023-NR077289 and RS-2024-00358448.
\vspace{-5pt}
\bibliographystyle{IEEEbib}
{
\small  
\setstretch{0.9}
\bibliography{strings,refs}
}
\end{document}